\documentclass[11pt]{article}             
\usepackage{osid}                         %

\usepackage{hyperref}
\usepackage{graphicx}
\usepackage{color}
\usepackage{amssymb}

\usepackage{hyperref}

\newcommand{\As}{\mathcal{A}}
\newcommand{\bm}[1]{\mbox{\boldmath $#1$}}

\def\bra#1{\langle #1 |}
\def\ket#1{| #1 \rangle}

\def\Ord{\mathrm{O}}
\renewcommand{\Re}{\mathrm{\,Re\,}}
\renewcommand{\Im}{\mathrm{\,Im\,}}

\newcommand{\beq}{\begin{equation}}
\newcommand{\eeq}{\end{equation}}
\newcommand{\barr}{\begin{eqnarray}}
\newcommand{\earr}{\end{eqnarray}}

\def\bra#1{\langle #1 |}
\def\ket#1{| #1 \rangle}

\title{All you ever wanted to know about the quantum Zeno effect in 70 minutes
}
\author{Saverio Pascazio\\{\footnotesize\it Dipartimento di Fisica and MECENAS,
Universit\`a di Bari, I-70126 Bari, Italy \\
\& INFN, Sezione di Bari, I-70126 Bari, Italy \\ email:  saverio.pascazio@ba.infn.it}
}

\begin{document}

\maketitle
\begin{abstract}
This is a primer on the quantum Zeno effect, addressed to students and researchers with no previous knowledge on the subject. The prerequisites are the Schr\"odinger equation and the von Neumann notion of projective measurement. 
\end{abstract}

\date{\today}


\section{Introduction and motivation}

\begin{figure}[t]
\begin{center}
\includegraphics[width=8cm]{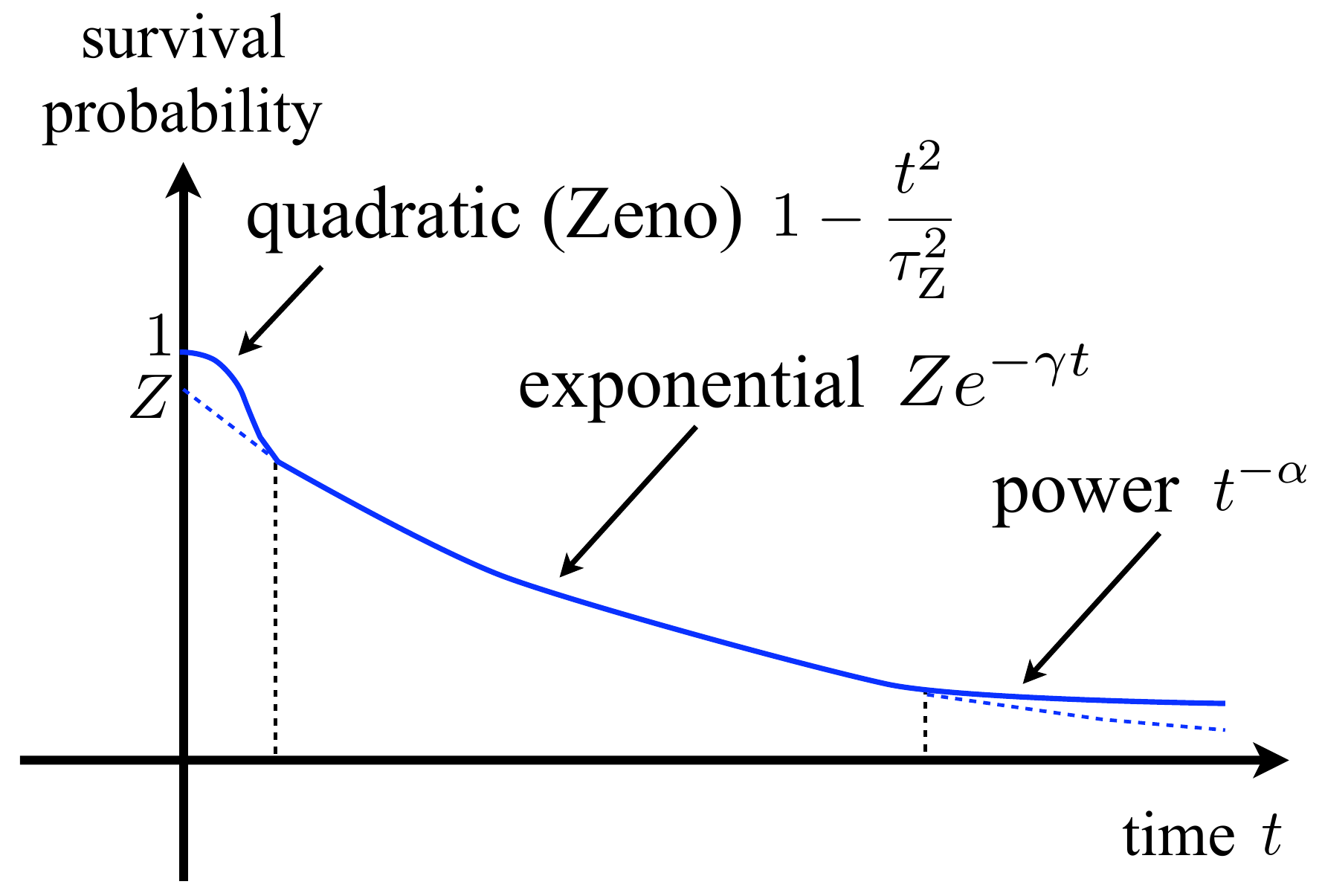}
\end{center}
\caption{Survival probability of a decaying quantum system. The initial Zeno region is followed by an exponential decay and finally superseded by a power law. Notice that the extrapolation of the exponential law back to $t=0$ yields a value $Z$ that is in general $\neq 1$.}
\label{fig:genevol}
\end{figure}

The evolution of an unstable quantum system is characterized by three distinct regimes \cite{strev,zenoreview}: a short-time region, where the decay is quadratic, an intermediate region, during which the exponential law sets in, and a long-time region, governed by a power law. 
A sketch (not in scale!) of such an evolution is given in Fig.\  \ref{fig:genevol}.

Unlike in classical (statistical) mechanics, where a decaying system is treated heuristically and the exponential decay law is easily obtained, the quantum analysis turns out to be involved and sometimes difficult to follow, even for experienced physicists. Scrutiny of the quantum evolution, governed by the Schr\"odinger equation, unveils the presence of an unavoidable quadratic region at short (sometimes \emph{very} short) times. This region was baptized ``Zeno" by Misra and Sudarshan \cite{Misra} in 1977. The classical allusion to the sophist philosopher is due to an intriguing application: if one frequently interrogates the system, checking whether it is still in its initial state, one can slow down (and eventually stop) its evolution \cite{strev,zenoreview,KK}. This is similar to Zeno's arrow, that would not reach its target if observed at a given position \cite{aristotle}.

The purpose of this note is to give an introduction to this topic, addressed to students, young researchers and physicists with no previous knowledge on the subject.
This is the summary of a 70 minute lecture \cite{video} delivered in Toru\'n, Poland, on June 21th, 2012, during the 44th Symposium on Mathematical Physics on ``New Developments in the Theory of Open Quantum Systems". The audience provided an excellent arena to test the pedagogical aspects of the lecture and helped me understand which facets of the problem are more difficult to grasp. I can only hope that I succeeded in making my presentation as palatable as possible. 
Some of the examples investigated here have been presented elsewhere \cite{zenoreview,waseda01}. I do not aim at novelty, but rather at clarity, sometimes at the expenses of rigor.

These notes are a somewhat more detailed version of the lecture. The level of the presentation will be kept as elementary as possible. The reader is invited to perform \emph{all} calculations.

In Sec.\ \ref{sec-prelnot} we review the main features of the quantum evolution law. These are straightforward consequences of the Schr\"odinger equation. We introduce the quantum Zeno effect in Sec.\ \ref{sec-dpw}. As anticipated, it is a very general, unavoidable by-product of the quantal dynamics.
We then clarify these general aspects by looking at the simplest non-trivial quantum mechanical example (a two-level system) in Sec.\ \ref{sec-2levels}
We briefly comment on the physical and mathematical origin of the quantum Zeno region in Sec.\ \ref{sec-comments} and on the ``meaning" of a von Neumann projective measurement in Sec.\ \ref{sec-QZEnH}

The general analysis of the Zeno effect is disguisingly simple. In Sec.\ \ref{sec-unstQZE} we turn to genuinely unstable systems, that require a quantum field theoretical description, and derive a closed expression for the survival amplitude. The analysis makes use of an analytic continuation in the complex energy plane. Before embarking in this adventure, we remind in Sec.\ \ref{sec-maths} how to perform analytic continuations to the second Riemann sheet, in presence of a cut singularity.
The analytic continuation of the propagator is done in Sec.\ \ref{sec-ancontprop}. We conclude (and apologize) in Sec.\ \ref{sec-concl}

\section{The quantum mechanical evolution}
\label{sec-prelnot}

\subsection{Evolution with Hermitian Hamiltonian}
\label{sec-HH}

\begin{figure}
\begin{center}
\includegraphics[width=8cm]{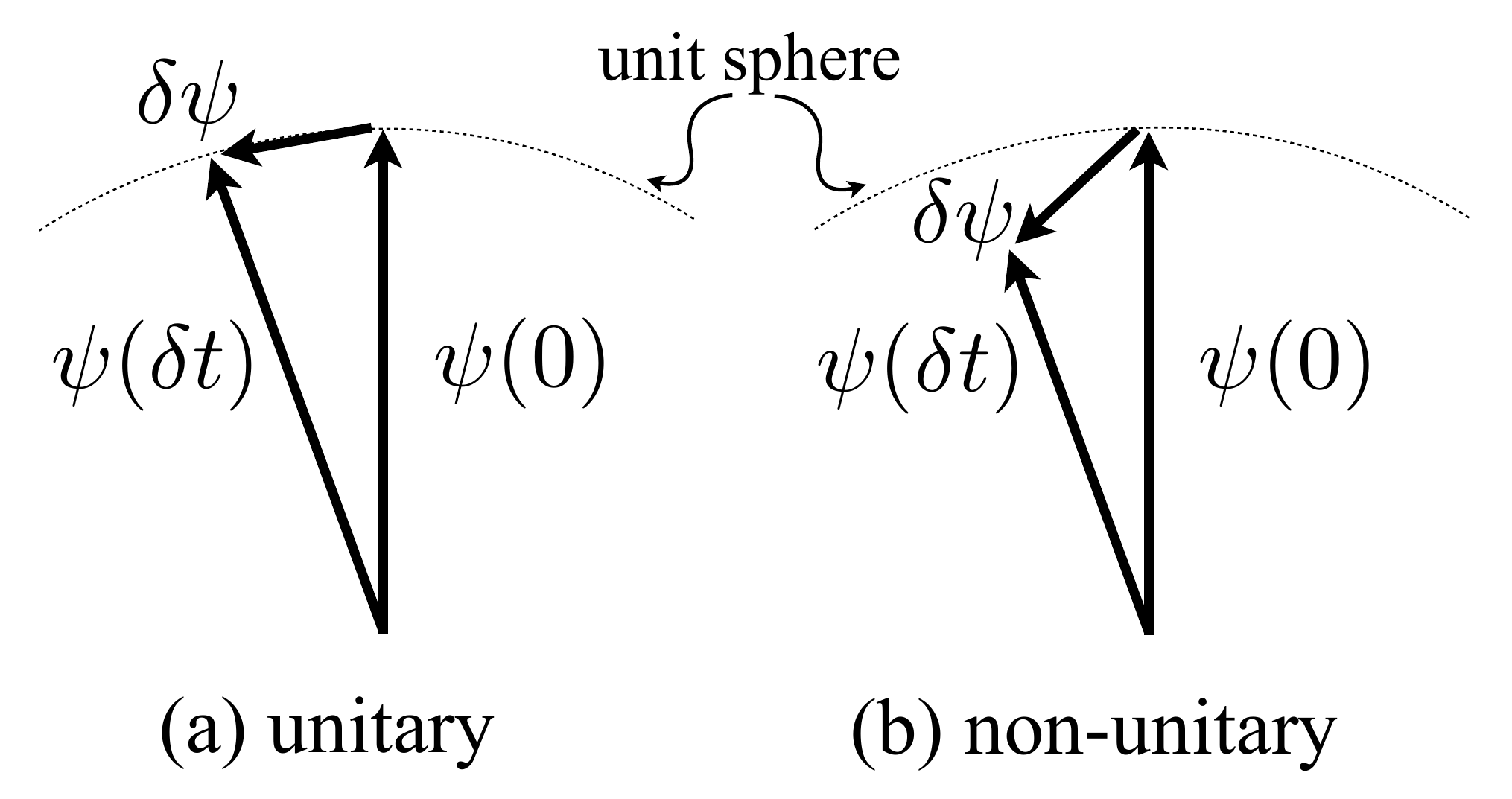}
\end{center}
\caption{(a) Unitary evolution engendered by a Hermitian Hamiltonian. The evolution takes place on the unit sphere: $||\psi (\delta t)||=||\psi (0)||=1$.
(b) Non-unitary evolution engendered by a non-Hermitian Hamiltonian. The tip of the state vector can leave the unit sphere (and enter the unit ball): $||\psi (\delta t)||\leq ||\psi (0)||=1$. In both cases, $\delta \psi$ is linear in $\delta t$.
}
\label{fig:unitsphere}
\end{figure}

We start off by scrutinizing the quantum-mechanical evolution law, focusing on its short-time features.
Let $H$ be the Hamiltonian of a quantum system and $\ket{\psi_0} = \ket{\psi (t=0)} $ its initial state.
We shall set henceforth $\hbar=1$ and assume that all functions to be dealt with are sufficiently regular to admit series expansions.
We shall focus on the ``survival" amplitude $\As$ and probability $p$ that the system has survived in its initial state $\ket{\psi_0}$ at time $t$:
\barr
\As (t) &=& \langle \psi_0|\psi (t)\rangle = \langle \psi_0|e^{-iHt}|\psi_0\rangle , \label{eq:unoa}\\
p(t) &=& |\As (t)|^2 =|\langle \psi_0|e^{-iHt}|\psi_0\rangle |^2.
\label{eq:unob}
\earr
Let the system evolve for a short time $\delta t$. The Schr\"odinger equation yields
\barr
|\psi(\delta t)\rangle = e^{-iH\delta t}|\psi_0\rangle & = & |\psi_0\rangle -iH |\psi_0\rangle \delta t -\frac{1}{2} H^2 |\psi_0\rangle (\delta t)^2 + \Ord((\delta t)^3) \nonumber \\
&\equiv& |\psi_0 \rangle + |\delta \psi \rangle .
\label{eq:psidt}
\earr
The short-time expansion (\ref{eq:psidt}) yields
\barr
\As(\delta t) & = & 1 - i  \langle H \rangle_0 \delta t - \frac{1}{2} \langle H^2 \rangle_0 (\delta t)^2 ,
\label{eq:linas}
 \\
p(\delta t) & = & 1 - \frac{(\delta t)^2}{\tau_{\mathrm{Z}}^2} + \Ord((\delta t)^4) ,
\label{eq:quadratic}
\earr
where $ \langle  \cdots \rangle_0 \equiv \langle \psi_0|\cdots |\psi_0\rangle$ and
\beq
\tau_{\mathrm{Z}}^{-2} \equiv \langle H^2\rangle_0 - \langle H\rangle_0^2 ,
\label{eq:tauz}
\eeq
is the Zeno time \cite{zenoreview}.
In deriving  (\ref{eq:quadratic}) from (\ref{eq:linas}) the Hermitianity of $H$, ensuring the reality of $\langle H \rangle_0$, played a primary role. Notice that according to (\ref{eq:linas}) the wave function evolves linearly away from the initial state, but the survival probability (of remaining in the initial state) evolves \emph{quadratically} away from 1, due to (\ref{eq:quadratic}). Recall that due to the unitarity of the evolution, wave functions are always normalized to unity: $||\psi(t)||=||\psi (0)||=1, \forall t$: the tip of the state vector never leaves the unit sphere. The features of the short time evolution are pictorially displayed in Fig.\  \ref{fig:unitsphere}(a).

\subsection{Evolution with non-Hermitian Hamiltonian}
\label{sec-nHH}

Let us add a non-Hermitian part to the Hamiltonian:
\beq
H' = H - iV,
\label{eq:HiV}
\eeq
where $V>0$ is a real ``optical" potential (taken to be independent of all dynamical variables---such as position---for simplicity). Optical potentials were frequently used by the founding fathers of nuclear physics, who introduced them in order to describe the coherent scattering of slow neutrons traveling through matter \cite{opticalpotential}\footnote{The term ``optical" is due to the analogy with the interaction of light with a medium that is both refractive and absorptive. Such an interaction can be analyzed by introducing a complex refractive index. Analogously, the scattering and absorption of nucleons by nuclei can be treated by introducing effective neutron-nucleus interaction potentials and by averaging such effective potentials over many nuclei in order to obtains the neutron-matter (complex) optical potential. A consistent expression of $V$ was first derived by Fermi and Zinn \cite{fermizinn}.}.

The new survival amplitude and probability read
\barr
\As' (t) &=& \langle \psi_0|\psi (t)\rangle = e^{-Vt} \langle \psi_0|e^{-iHt}|\psi_0\rangle , \label{eq:unoaprime}\\
p'(t) &=&  e^{-2Vt} |\langle \psi_0|e^{-iHt}|\psi_0\rangle |^2.
\label{eq:unobprime}
\earr
A short-time expansion yields a \emph{linear} behavior both for amplitude and probability
\barr
\As' (\delta t) &=&
 1 - (V + i  \langle H \rangle_0) \delta t - \frac{1}{2} (\langle H^2 \rangle_0 - V^2 - 2iV \langle H \rangle_0 ) (\delta t)^2 + \Ord((\delta t)^3), 
\nonumber \\
\label{eq:linear1} \\
p'(\delta t)  &=&  1 - 2V\delta t + \Ord((\delta t)^2).
\label{eq:linear2}
\earr
Optical potentials ``eat up" probability and account for decay channels. See Fig.\   \ref{fig:unitsphere}(b). The tip of the state vector can leave the unit sphere and enter the unit ball: $||\psi (t)|| \leq ||\psi (0)||=1$. [It would leave the unit ball if the optical potential $-iV$ in
(\ref{eq:HiV}) had the opposite sign.]

In physics, one tends to regards property (\ref{eq:quadratic}) as more ``fundamental", as it ensues from the Hermitianity of the Hamiltonian and the unitarity of the evolution, that are regarded as very general principles.
Yet optical potentials have their own charm and play an important role in effective descriptions of decaying and dissipative systems. Nowadays they have been superseded by the rigorous mathematical framework of
Gorini, Kossakowski, Sudarshan and Lindblad \cite{GKSL} that describes the physics of quantum dissipative systems
\cite{Alicki,Weiss,Breuer}.

It is also worth noticing that the exponential law in a quantum context is always the consequence of approximations of some sort. Examples of such approximations can be a macroscopic limit \cite{colhepp} or the intervention of an external apparatus, governed by classical laws, that interacts with the system investigated \cite{measdiff}.

\subsection{Interaction Hamiltonian}
\label{sec-IH}

If the Hamiltonian is composed of a free and an interaction parts
\beq
H = H_0 + H_{\mathrm{int}} 
\eeq
we can obtain an interesting expression, that sheds light on the meaning of the Zeno time.
Let $\ket{\psi_n}$ be the eigenstates of the free Hamiltonian, that form a complete set
\beq
H_0\ket{\psi_n} = \omega_n\ket{\psi_n}.
\eeq
We require that the initial state be an eigenstate of the free Hamiltonian and
(as it is customary in quantum field theory) that the interaction be off-diagonal:
\beq
H_0\ket{\psi_0} = \omega_0\ket{\psi_0},    \qquad \langle H_{\mathrm{int}} \rangle_0=0. \label{eq:Hdiv}
\eeq
In this interesting case the Zeno time reads
\beq
\tau_{\mathrm{Z}}^{-2} = \langle H_{\mathrm{int}}^2 \rangle_0 = \sum_n
\langle\psi_0| H_{\rm int}|\psi_n\rangle
\langle\psi_n| H_{\rm int}|\psi_0\rangle
\label{eq:tzoff}
\eeq
and depends only on the interaction Hamiltonian.

Formula (\ref{eq:tzoff}) should be compared to the Fermi ``golden rule" \cite{Fermi}\footnote{Fermi considered expression (\ref{eq:goldenrule}) the \emph{second} golden rule. If you are curious about the first one, see pages 136 and 148 of {\it Nuclear Physics} \cite{Fermi}.}, yielding the inverse lifetime $\gamma$ of a decaying quantum system:
\beq
\gamma = 2 \pi \sum_{f}
\left|\langle \psi_f| H_{\rm int}|\psi_0\rangle\right|^2
\delta(\omega_f-\omega_0),
\label{eq:goldenrule}
\eeq
where the summation (integral) is over the final states and the continuum limit is implied.

\begin{figure}
\begin{center}
\includegraphics[width=8cm]{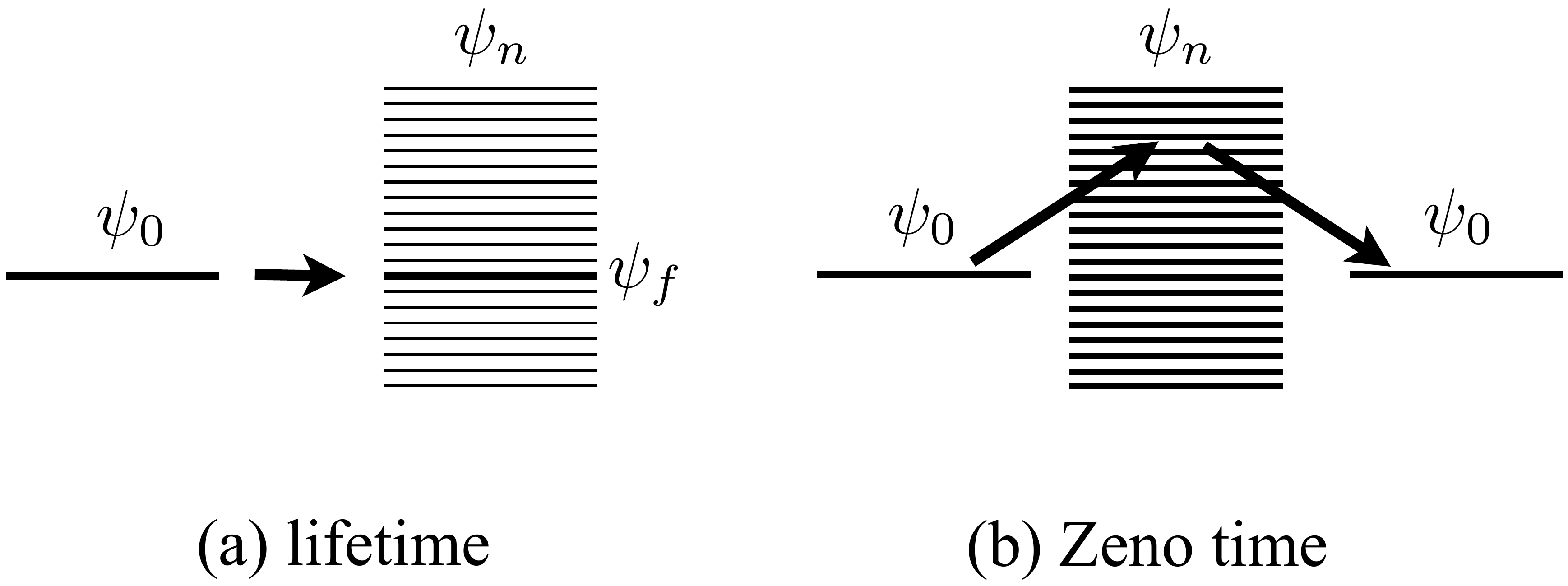}
\end{center}
\caption{(a) The lifetime $\gamma$ in Eq.\ (\ref{eq:goldenrule}) contains only ``on-shell" contributions: the delta function entails energy conservation $\omega_f=\omega_0$; $\psi_f$ is in general (very) degenerate (think of an atom in an $S$-wave emitting a photon: there is a $4\pi$ degeneracy in the direction of emission). (b) The Zeno time $\tau_{\mathrm{Z}}$ in Eq.\ (\ref{eq:tzoff}) explores the whole Hilbert space. }
\label{fig:zenotime}
\end{figure}

One comment. While (\ref{eq:goldenrule}) contains only ``on-shell" contributions (because the delta function ensures energy conservation), the expression (\ref{eq:tzoff}) explores the \emph{whole} Hilbert space.
See Fig.\ \ref{fig:zenotime}.

\section{Quantum Zeno effect}
\label{sec-dpw}

The most familiar formulation of the QZE makes use of Von
Neumann measurements, represented by one-dimensional projectors.
Perform $N$ measurements at time intervals $\tau=t/N$, in order to check whether the system is
still in its initial state $\ket{\psi_0}$. After each measurement the system's state is ``projected" back onto its initial state
$\ket{\psi_0}$ and the evolution starts anew according to Schr\"odinger's equation with initial condition $\ket{\psi_0}$. [The system can also be projected onto an orthogonal state $\ket{\psi_0^\perp}$, with (quadratic) probability $1- p(\tau) = \tau^2/\tau_{\mathrm{Z}}^2$, according to Eq.\ (\ref{eq:quadratic}). As $\tau=\Ord(1/N)$, such an event becomes increasingly unlikely as $N$ increases.]

The survival probability
$p^{(N)}(t)$ at the final time $t=N \tau$  reads
\barr
p^{(N)}(t)&=&p(\tau)^N = p(t/N)^N\nonumber\\
&\simeq& \left[ 1 - (t/N\tau_{\rm Z})^2 \right]^N \stackrel{N \;
{\rm large}}{\longrightarrow} \exp(-t^2/N\tau_{\rm Z}^2)
\stackrel{N \rightarrow\infty}{\longrightarrow} 1 ,
 \label{eq:survN}
\earr
where we made use of Eq.\ (\ref{eq:quadratic}). For large $N$ the quantum mechanical evolution is slowed down and in the $N\to\infty$ limit (infinitely frequent measurements) it is halted, so that the state of the system is ``frozen" in its initial state. This is the QZE. It is a consequence of the short-time behavior (\ref{eq:quadratic}).

\begin{figure}[t]
\begin{center}
\includegraphics[width=8cm]{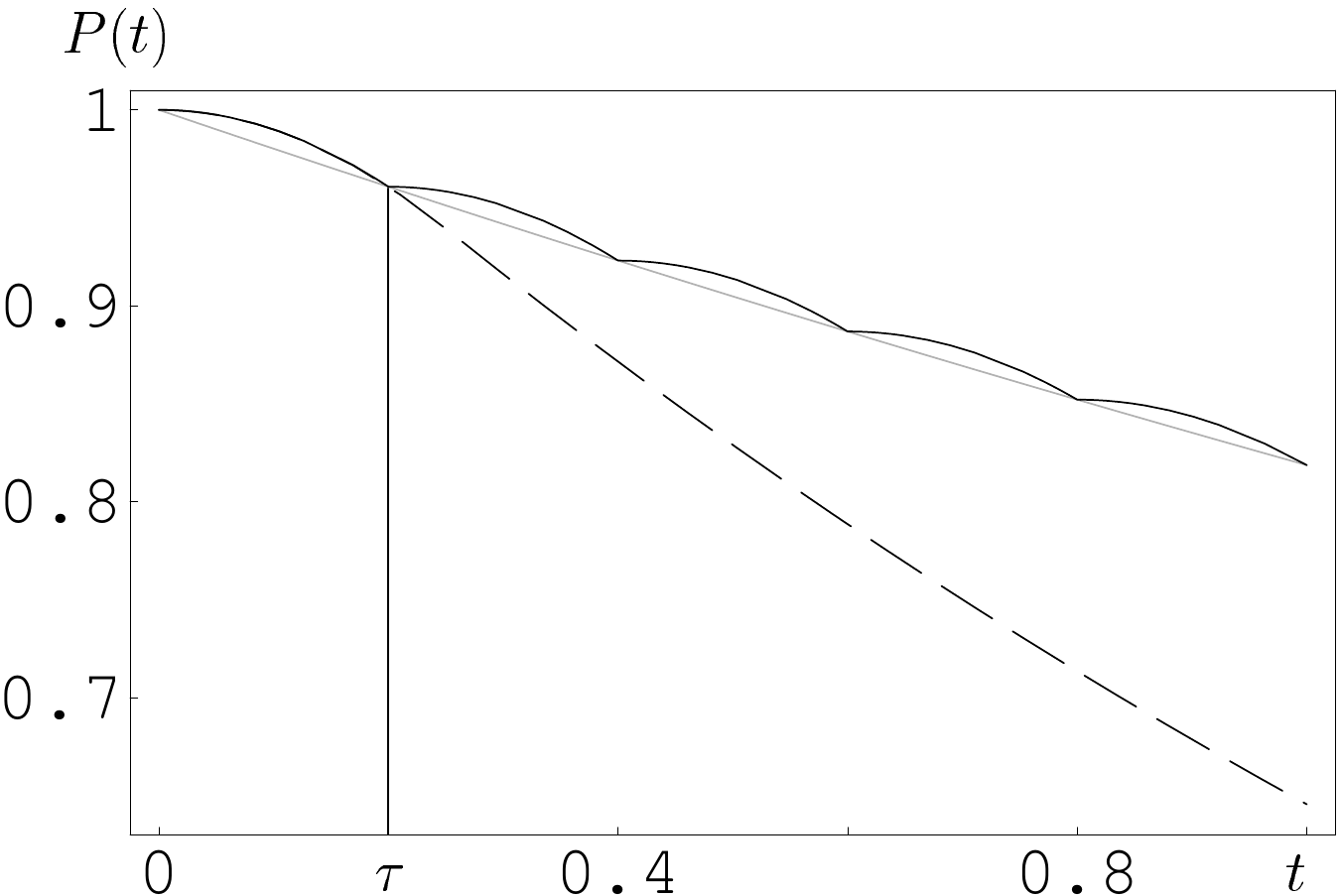}
\end{center}
\caption{Quantum Zeno effect  for $N=5$ ``pulsed" Von Neumann measurements.
The dashed (full) line is the survival probability without (with)
measurements. The gray line is the interpolating exponential
(\ref{eq:survN0}). As $N$ increases, $p^{(N)}(t) \to 1$ uniformly in $[0,t]$.
The units on the abscissae are arbitrarily chosen for illustrative purposes.}
\label{fig:zenoevol}
\end{figure}

Observe that the survival probability after $N$ pulsed measurements ($t=N\tau$) is
interpolated by an exponential law \cite{heraclitus}
\beq
p^{(N)}(t)=p(\tau)^N=\exp(N\log p(\tau))=
\exp(-\gamma_{\mathrm{eff}}(\tau) t) ,
\label{eq:survN0}
\eeq
with an effective decay rate
\beq
\gamma_{\mathrm{eff}}(\tau) \equiv -\frac{1}{\tau}\log p(\tau)  . \label{eq:gammaeffdef}
\eeq
For $\tau\to 0 $ ($N \to \infty$) one gets from (\ref{eq:quadratic}) $p(\tau) \simeq
\exp (-\tau^2/\tau_{\mathrm{Z}}^2)$, so that
\beq
\gamma_{\mathrm{eff}}(\tau)\simeq \tau/\tau_{\mathrm{Z}}^2,
\qquad \tau\to 0.
\label{eq:lingammaeff}
\eeq
The Zeno evolution for ``pulsed" Von Neumann measurements is pictorially
represented in Figure \ref{fig:zenoevol}.

\section{The simplest non-trivial quantum mechanical example: the two-level system}
\label{sec-2levels}

Consider a two-level system undergoing Rabi
oscillations. This is the simplest nontrivial quantum mechanical
example, for it involves $2\times 2$ matrices and very simple
algebra. One can think of an atom shined by a driving laser field whose
frequency resonates with one of the atomic transitions, or a
neutron spin in a magnetic field. The (interaction) Hamiltonian reads
\beq
H=
H_{\mathrm{int}} = \Omega \sigma_1 = \Omega ( |+\rangle \langle -| +
|-\rangle \langle +|) = \pmatrix{0 & \Omega \cr \Omega & 0},
\label{hamrabi}
\eeq
where $\Omega$ is a real number, $\sigma_j \; (j=1,2,3)$ the Pauli
matrices and
\beq
|+\rangle = \pmatrix{1 \cr 0} , \quad |-\rangle = \pmatrix{0 \cr
1}
\label{+-}
\eeq
are eigenstates of $\sigma_3$. We are neglecting the energy
difference between the two states $|\pm \rangle$. Let the initial state be
\beq
|\psi_0\rangle = |+\rangle = \pmatrix{1 \cr 0},
\label{inrabi}
\eeq
so that the evolution yields
\beq
|\psi (t)\rangle = e^{-iH_{\mathrm{int}}t}|\psi_0\rangle = \cos (\Omega t)
|+\rangle - i \sin (\Omega t) |-\rangle = \pmatrix{\cos \Omega t
\cr - i \sin \Omega t}.
\label{rabit}
\eeq
The survival amplitude (\ref{eq:unoa}) and probability (\ref{eq:unob}) and the Zeno time
(\ref{eq:tauz}) or (\ref{eq:tzoff}) read
\barr
\As (t) & = & \cos \Omega t,
\label{eq:Arabi} \\
p(t) & = & \cos^2 \Omega t,
\label{eq:Prabi}\\
\tau_{\rm Z} & = & \Omega^{-1},
\label{eq:tauzrabi}
\earr
respectively.
The effective decay rate (\ref{eq:gammaeffdef}) if $N$ measurements are performed in time $t$ reads
\beq
\gamma_{\mathrm{eff}}(\tau) = \tau\Omega^2.
\label{eq:gammaeff2lev}
\eeq
In this simple case, Eq.\ (\ref{eq:lingammaeff}) is exact (and not simply an approximation for short $\tau$).
Look again at Figure \ref{fig:zenoevol}.

\section{Comments.}
\label{sec-comments}

\begin{figure}[t]
\begin{center}
\includegraphics[width=6cm]{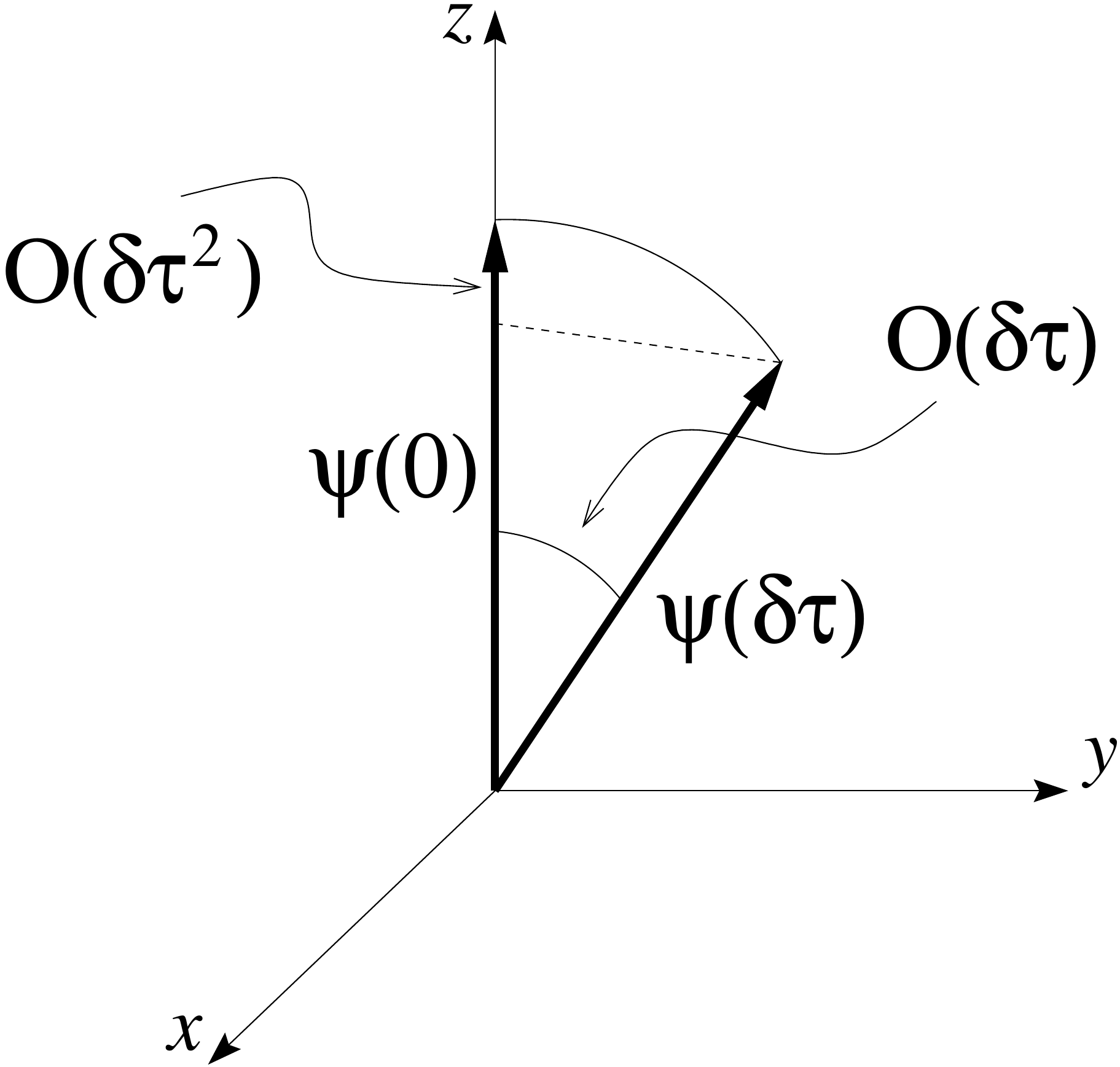}
\end{center}
\caption{Short-time evolution of phase and probability: $\delta \tau \sim 1/N$.}
\label{fig:phprob}
\end{figure}

At the end of the day, the QZE is ascribable to the following mathematical properties of the Schr\"o\-din\-ger equation: in a short time $\delta\tau (\sim 1/N)$, the phase of the wave function evolves like $\Ord (\delta\tau)$, while the probability changes by $\Ord (\delta\tau^2)$, so that
\beq
P^{(N)}(t) \simeq \left[ 1 - \Ord (1/N^2)\right]^N \stackrel{N
\rightarrow \infty}{\longrightarrow}1.
 \label{eq:survN3}
\eeq
Stated differently, the projection onto the inital state ``slowly" evolves away from unity. This is sketched in Fig.\ \ref{fig:phprob} and is a very general feature of the Schr\"odinger equation, as well as of other ``fundamental" evolution equations in physics\footnote{Such as the Maxwell equations and (super-)renormalizable quantum field theories.}. Equations that do not have this feature (e.g.\ dissipative equations) tend to be regarded as less fundamental, the consequence of approximations of some sort.

\section{Unraveling a von Neumann measurement}
\label{sec-QZEnH}

What is a (von Neumann \cite{von}) measurement? This is a difficult question, that has been debated for decades and is still a subject of controversy \cite{pollQM}. The mathematical answer is clear, the physical one is not.
The evolution due to a measurement process is non-unitary and many reasons lead many physicists (including myself) to think that a von Neumann \emph{projection} is but an effective description of a quantum measurement process. Stated differently, von Neumann's projectors are a short-hand notation: they summarize the complicated processes that take place in the macroscopic apparata that perform the measurement and are placed in (macroscopic) regions of space-time \cite{WZ83,NP93}.

When one deals with the quantum Zeno effect, the situation gets even worse. In general, in order to interact with different ``projectors", the physical system of interest must move, traversing mesoscopic or macroscopic regions of space\footnote{There are situations where the system need not move between measurements, but they are rare, and presuppose the existence of a control mechanism that keeps at a given place the physical system undergoing the measurement. An example is an atom in a given position that is shined by a laser: by observing the photons that are scattered/emitted, one can infer which atomic level is populated.}. This dynamics is neglected in most analyses of the QZE: in the elapse of time between two subsequent projections, the system evolves under the action of a Hamiltonian $H$ that does not account for its movement from the region of space where one projection occurs to the (macroscopically) different region of space where the following projection will take place. Think of the example in Sec.\ \ref{sec-2levels}: everything was neglected but the two-level structure of the system. The physics behind the measurement process is dismissed altogether in a single sentence after Eq.\ (\ref{eq:tauzrabi})! We are so accustomed at computing projections that we do not even think about the underlying physical processes anymore.

We shall henceforth neglect all these problems and act pragmatically. In this section we forget philosophical standpoints and personal taste, and endeavor to give a heuristic description of a quantum measurement, by proposing an effective model for the measuring ``apparatus". Clearly, we are not even hoping of contributing to solving the mistery behind a quantum measurement.

\subsection{Mimicking the projection with a non-Hermitian Hamiltonian}
\label{sec-QZEnH1}

Let us show that the action of a measuring apparatus (performing the Von Neumann measurement) can
be mimicked by a non-Hermitian Hamiltonian. Consider the Hamiltonian (notation as in Sec.\ \ref{sec-2levels})
\beq
H_{\mathrm{int}} = \pmatrix{0 & \Omega \cr \Omega & -i2V} = -iV{\bf 1} +
\bm h \cdot \bm\sigma ,
\label{iV}  \qquad \bm h =(\Omega,0,iV)^T ,
\eeq
that yields Rabi oscillations of frequency $\Omega$, but at the
same time absorbs away the $|-\rangle$ component of the state vector, performing in this way a ``measurement." $H$ is non-Hermitian, therefore probabilities are not conserved: we are focusing our attention only on the $|+\rangle$ component.
State $|-\rangle$ can be viewed as a ``decay channel", according to the discussion in Sec.\ref{sec-nHH}.

\begin{figure}[t]
\begin{center}
\includegraphics[width=8cm]{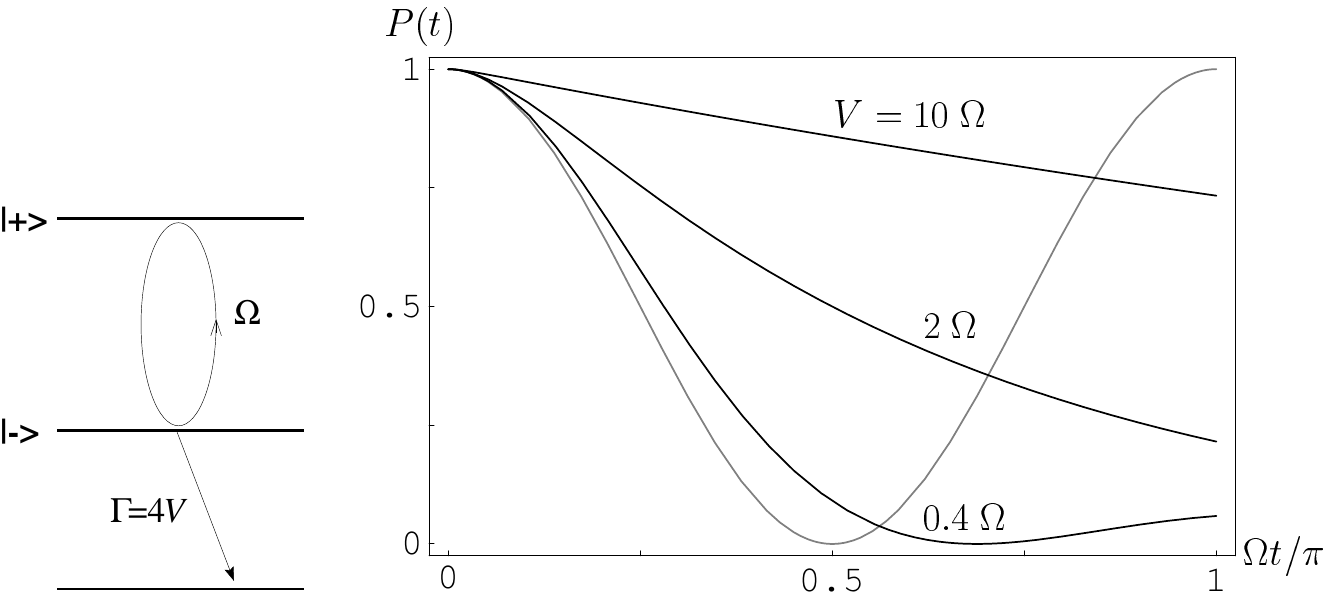}
\end{center}
\caption{Survival probability for a system undergoing Rabi
oscillations in presence of absorption ($V=0.4, 2, 10 \Omega$).
The gray line is the undisturbed evolution ($V=0$). }
\label{fig:zenoiV}
\end{figure}

Elementary algebra [and properties of SU(2)] yields 
\beq
e^{-iH_{\mathrm{int}}t} = e^{-Vt} \left[ \cosh (ht) -i \frac{\bm h\cdot
\bm \sigma}{h}\sinh (ht) \right],
\label{eiVt}
\eeq
where $h=\sqrt{V^2-\Omega^2}$ and we supposed $V\gg\Omega$ (this hypothesis is not vital, but makes the measurement ``fast" and therefore effective). The survival amplitude in the initial state (\ref{inrabi}) reads
\barr
\As (t) &=& \langle \psi_0 |e^{-iH_{\mathrm{int}}t}|\psi_0\rangle \nonumber\\
&=& e^{-Vt} \left[ \cosh (ht) +
\frac{V}{h}\sinh (ht) \right] \nonumber \\
&=& \frac{1}{2} \left( 1 + \frac{V}{h} \right)
e^{-(V-h)t} + \frac{1}{2} \left( 1 - \frac{V}{h} \right)
e^{-(V+h)t}.
\label{eq:survamplV}
\earr
Notice the presence of a slow and a fast decay. The survival
probability $P(t)=|\As(t)|^2$ is shown in Fig.\ \ref{fig:zenoiV}
for $V=0.4,2,10 \Omega$.

As expected, probability is (exponentially) absorbed away as $t \to \infty$. Moreover, for large $V$, by expanding in the small parameter $\Omega/V$, one finds 
\beq
P(t)\simeq\left(1+\frac{\Omega^2}{2V}\right)
\exp\left(-\frac{\Omega^2}{V} t \right),
\label{eq:largeV}
\eeq
where the wrong normalization at $t=0$ is an artifact of the approximation (the decay is always quadratic at short times and the above expansion becomes accurate very quickly, on a time scale of order $V^{-1}$). The effective decay rate $\gamma_{\rm eff}(V)=\Omega^2/V$ is counterintuitive. Try and show the left panel in Fig.\ 
\ref{fig:zenoiV} to a friend or a colleague of yours, who has no familiarity with the QZE, and ask the following question: what happens if one initially populates state $|+\rangle$ and increases the decay rate $V$ out of state $|-\rangle$? Chances are that your friend/colleague will reply: state $|+\rangle$ will be depleted faster. Not so: $V$ appears in the \emph{denominator} of the exponent in Eq.\ (\ref{eq:largeV}). Now show your friend the right panel in Fig.\ \ref{fig:zenoiV}. The effective lifetime becomes larger as $V$ increases, eventually halting the ``decay" (absorption) of the initial state in the $V \to \infty$ limit. A larger $V$ entails a more ``effective" measurement of the initial state.
This is an interesting example of QZE.

The global process described here can be viewed as a ``continuous" (negative result) measurement performed on the initial state $\ket{+}$. State $\ket{-}$ is continuously monitored with a response time $1/V$: as soon as it becomes populated, it is detected within a time $1/V$. The ``strength"
$V$ of the observation can be compared to the frequency $\tau^{-1}= (t/N)^{-1}$ of measurements in the ``pulsed" formulation of Sec.\ \ref{sec-dpw}. Indeed, for large values of $V$ one gets from Eq.\ (\ref{eq:largeV})
\beq
\gamma_{\rm eff}(V) = \frac{\Omega^2}{V}=\frac{1}{\tau_{\rm Z}^2 V}, 
\label{eq:effgammaG}
\eeq
which, compared with Eq.\ (\ref{eq:lingammaeff}), yields a cute relation between continuous and pulsed measurements
\cite{Schulman98}
\beq
V\simeq 1/\tau.
\label{eq:contvspuls}
\eeq

\subsection{Interaction with an external field yields a non-Hermitian Hamiltonian}
\label{sec-QZEnH2}

We now show that the non-Hermitian Hamiltonian (\ref{iV}) can be
obtained by considering the evolution engendered by a Hermitian
Hamiltonian acting on a larger Hilbert space and then restricting
the attention to the subspace spanned by $\{\ket{+}, \ket{-}\}$.
Let
\beq
H= \Omega(\ket{+}\bra{-}+\ket{-}\bra{+}) +\int d\omega\;\omega
\ket{\omega} \bra{\omega} +\sqrt{\frac{\Gamma}{2\pi}}\int
d\omega\;( \ket{-} \bra{\omega}+ \ket{\omega} \bra{-}) ,
\label{eq:hamflat}
\eeq
that describes a two-level system coupled to a one-dimensional massless boson field in
the rotating-wave approximation. Notice that the coupling is ``flat": the two-level system couples to all frequencies in the same way: this enables us to pull out of the last integral a coupling constant $\sqrt{\Gamma}$ that is equal for all frequencies. The state of the system at time $t$ can be written as
\beq
\ket{\psi(t)}=x(t)\ket{+}+y(t)\ket{-}+\int d\omega\;
z(\omega,t)\ket{\omega}
\label{eq:stateflat}
\eeq
and the Schr\"odinger equation reads
\barr
i\dot x(t)&=& \Omega y(t), \nonumber\\
i\dot y(t)&=& \Omega x(t) +\sqrt{\frac{\Gamma}{2\pi}}\int
d\omega\;
z(\omega,t), \label{eq:equsflat}\\
i\dot z(\omega,t)&=&\omega
z(\omega,t)+\sqrt{\frac{\Gamma}{2\pi}}y(t). \nonumber
\earr
By using the initial condition $x(0)=1$ and $y(0)=z(\omega,0)=0$
one obtains
\beq
z(\omega,t)=-i\sqrt{\frac{\Gamma}{2\pi}}\int_0^t d\tau\;
e^{-i\omega(t-\tau)}y(\tau)
\label{eq:zflat}
\eeq
and
\beq
i\dot y(t)= \Omega x(t) -i\frac{\Gamma}{2\pi}\int d\omega\int_0^t
d\tau\; e^{-i\omega(t-\tau)}y(\tau)=\Omega x(t) - i
\frac{\Gamma}{2} y(t).
\label{eq:yequflat}
\eeq
Observe that in order to obtain this result the integral over $\omega$ has to be extended over the whole real line (from $-\infty$ to $+\infty$).  Also, $\int_0^t \delta(t-\tau) d \tau = 1/2$. 

The only remnant of the coupling of the qubit to the continuum of levels is the appearance of the imaginary frequency $-i\Gamma/2$. This is ascribable to the afore-mentioned ``flatness" of the continuum [there is no form factor or frequency cutoff in the interaction term of Eq.\ (\ref{eq:hamflat})], which yields a purely exponential (Markovian) decay of $y(t)$.

In conclusion, $z(\omega,t)$ drops out of the first two equations (\ref{eq:equsflat}), that now describe the (reduced) dynamics in the subspace spanned by $\ket{+}$ and $\ket{-}$:
\barr
i\dot x(t)&=& \Omega y(t), \nonumber\\
i\dot y(t)&=& -i \frac{\Gamma}{2} y + \Omega x(t).
\label{eq:subdynflat}
\earr
Of course, this dynamics is not unitary, for probability flows out of the subspace, and is generated by the non-Hermitian Hamiltonian
\beq
H=\Omega(\ket{+}\bra{-}+\ket{-}\bra{+})-i\frac{\Gamma}{2}
\ket{-}\bra{-}= \pmatrix{0 & \Omega \cr \Omega & -i\Gamma /2} .
\label{eq:nonhermham}
\eeq
This Hamiltonian is the same as (\ref{iV}) when one sets $\Gamma=4V$. QZE is obtained by increasing $\Gamma$: a larger coupling to the environment leads to a more effective ``continuous" observation on the system (quicker response of the measuring apparatus), and as a consequence to slower decay (QZE).
Try and ask the same tricky question mentioned after Eq.\ (\ref{eq:largeV}) to another friend/colleague. 
Rather than the left panel in Fig.\ \ref{fig:zenoiV}, draw a figure in which level $\ket{-}$ decays to a photon field, and increase the coupling $\Gamma$ between them. 

We leave it to the reader to judge whether the analysis of the measurement process proposed in this section is more satisfactory than that outlined in Sec.\ \ref{sec-QZEnH1}. We generally tend to regard this description more ``complete" than that proposed in Sec.\ \ref{sec-QZEnH1}.
One should notice that in this section quantum field theory has sneaked into the picture: equation (\ref{eq:hamflat}) describes a quantum field.

\section[Genuine unstable systems]{Genuine unstable systems and field theory}
 \label{sec-unstQZE}

We shall now forget about quantum measurements and QZE and focus on the non-exponential features of decay. The arguments given in Sec.\ \ref{sec-HH} are very general and cannot be rejected: decay cannot be exponential at short times. However, it is of great interest to discuss this problem in a quantum field theoretical framework. This will help us focus on the important role played by the form factors of the interaction. 

We start by generalizing the two-level Hamiltonian (\ref{hamrabi}) to $N$ states $\ket{j}$ ($j=1,\dots,N$) with different energies
\beq
H_0 = \omega_0\ket{+}\bra{+}+\sum_{j=1}^N\omega_j \ket{j} \bra{j}
= \pmatrix{\omega_0 & 0 & \ldots & 0 \cr
      0  & \omega_1 & \ldots & 0 \cr
    \vdots  & \vdots & \vdots & \vdots \cr
      0  & 0 & \ldots & \omega_N} .
\label{eq:ham0N}
\eeq
and (real) couplings
\beq
H_{\mathrm{int}} = \sum_{j=1}^N g_j ( \ket{+} \bra{j}+\ket{j}
\bra{+}) = \pmatrix{0 & g_1 & \ldots & g_N \cr
   g_1  & 0 & \ldots & 0 \cr
    \vdots  & \vdots & \vdots & \vdots \cr
   g_N  & 0 & \ldots & 0}
\label{eq:hamN}
\eeq
In order to obtain a truly unstable system we need a continuous
spectrum, so we consider the continuum limit $\omega_j \to \omega, \ket{j} \to \sqrt{\delta \omega}\ket{\omega}, g_j \to \sqrt{\delta \omega} g(\omega)$, with $\delta \omega \to 0$,
\beq
H=H_0+H_{\mathrm{int}} = \omega_0\ket{+}\bra{+} +\int d\omega\;\omega
\ket{\omega} \bra{\omega} +\int d\omega\;g(\omega) ( \ket{+}
\bra{\omega}+ \ket{\omega} \bra{+}) .
\label{eq:contham}
\eeq
State $\ket{+}$ is normalizable, but states $\ket{\omega}$ are not:
\beq
\langle + \ket{+} = 1, \quad \langle \omega\ket{\omega'} = \delta (\omega-\omega'), \quad 
\langle + \ket{\omega} =0.
\label{eq:normalization}
\eeq
$\{\ket{+},\ket{\omega}\}$ is the eigenbasis of $H_0$ and is a resolution of the identity
\beq
\label{eq:idres}
\ket{+}\bra{+}+\int d\omega\;\ket{\omega}\bra{\omega}= 1.
\eeq
As before, we take as initial state $\ket{\psi_0}=\ket{+}$. The interaction of this state with the continuum of states $\ket{\omega}$ is responsible for its decay and depends on the \emph{form factor} $g(\omega)$. 
We assumed (with no loss of generality) $g(\omega)$ to be real.

It is worth stressing that the purpose of studying model (\ref{eq:contham}) is very different from the motivations that led us to analyze model (\ref{eq:hamflat}). In Sec.\ \ref{sec-QZEnH2} we were interested in the QZE on level $\ket{+}$ that arises when level $\ket{-}$ is ``measured", while in this section we focus on the deviations from exponential when level $\ket{+}$ is coupled to a continuum. There is no level $\ket{-}$ here\footnote{Although it would not be difficult to introduce it.}.

The Fourier-Laplace transform of the survival amplitude (\ref{eq:unoa}) for this model can be given a convenient analytic expression. The transform of the survival amplitude is the expectation value of the \textit{resolvent}
\beq
\label{eq:transf}
\As(E)=\int_0^{\infty}dt\;e^{iEt}\As(t)
=\bra{+}\int_0^{\infty}dt\;e^{iEt}e^{-iHt}\ket{+}
=\bra{+}\frac{i}{E-H}\ket{+}
\eeq
and is defined for $\Im E > 0$ (check!). By using twice the operator
identity
\beq
\label{eq:operid}
\frac{1}{E-H}=\frac{1}{E-H_0}+\frac{1}{E-H_0}H_{\mathrm{int}}\frac{1}{E-H}
\eeq
one obtains
\barr
\As(E)&=& \!\! \bra{+}\left[\frac{i}{E-H_0}+\frac{1}{E-H_0}H_{\mathrm{int}}\frac{i}{E-H_0} + \right. \nonumber \\
 & & \qquad \left. + \frac{1}{E-H_0}H_{\mathrm{int}}\frac{1}{E-H_0}H_{\mathrm{int}}\frac{i}{E-H}\right]\ket{+}
\nonumber\\
&=&\frac{i}{E-\omega_0} +\frac{1}{E-\omega_0}\int d\omega\;
\frac{\left|\bra{+}H_{\mathrm{int}}\ket{\omega}\right|^2}{E-\omega}
\;\As(E).
\label{eq:transf1}
\earr
In the above derivation we used the resolution (\ref{eq:idres}) of the identity and the fact that $H_{\mathrm{int}}$ is completely off-diagonal in the eigenbasis of $H_0$ [compare Eq.\ (\ref{eq:Hdiv})].
The advantage of looking at the Fourier-Laplace transform (\ref{eq:transf}) lies in the fact that Eq.\ (\ref{eq:transf1}) 
is algebraic and can be solved to yield
\beq
\label{eq:propag}
\As(E)=\frac{i}{E-\omega_0-\Sigma(E)},
\eeq
where the \emph{self-energy function} $\Sigma(E)$ is related to the form
factor $g(\omega)$ by a simple integration
\beq
\label{eq:selfen}
\Sigma(E)=\int d\omega\; \frac{\left|\bra{+}H_{\mathrm{int}}\ket{\omega}\right|^2}{E-\omega} =\int
d\omega\;\frac{g^2(\omega)}{E-\omega}  .
\eeq
Notice that the self-energy function is a ``small" quantity, being proportional to the square of the coupling between level $\ket{+}$ and the continuum. By inverting Eq.\ (\ref{eq:transf}) we finally get
\beq
\label{eq:antitran}
\As(t)=\int_{\rm B}\frac{dE}{2\pi}\;e^{-iEt}\As(E)
=\frac{i}{2\pi}\int_{\rm B}dE\;
\frac{e^{-iEt}}{E-\omega_0-\Sigma(E)},
\eeq
the Bromwich path B being a horizontal line $\Im E=$constant$>0$ in the half plane of convergence of the
Fourier-Laplace transform (upper half plane). This is the quantity we sought, expressed in terms of a quadrature.

So far, the analysis is general and valid for any state. We shall now consider the case of an unstable system. However, before doing so, we shall give some mathematical notions related to complex analysis and analytic continuation.

\section{Intermezzo: Analytic continuation on the second Riemann sheet}
 \label{sec-maths}

Consider the function
\beq
F(z) = \int_0^\infty dE \frac{f(E)}{E-z}
\label{eq:Fss}
\eeq
where $z=x+iy \in \mathbb{C}$, $f$ is a smooth function and $E$ a real variable.
$F$ is an analytic function in the complex $z$ plane, but has a (logarithmic) cut
for positive real $z$. We obtain
\barr
F(x\pm i0^+) &=& \int_0^\infty dE \frac{f(E)}{E-x \mp i0^+}
\nonumber \\
&=& \int_0^\infty dE f(E) \left( \frac{\cal P}{E-x} \pm i\pi \delta
(E-x) \right) ,
\label{eq:Fzeropm}
\earr
where ${\cal P}$ denotes principal value.
The discontinuity across the cut is therefore 
\barr
F(x+i0^+) - F(x-i0^+) &=&
 2 \pi i \int_0^\infty dE f(E) \delta (E-x) =
 2 \pi i f(x)   \nonumber \\
 & &  \qquad \qquad \qquad \qquad \quad (x>0).
\label{eq:Fzerop}
\earr
Clearly, in Eq.\  (\ref{eq:Fzerop}) the function $F(x\pm i0^+)$ is evaluated on the {\em
first} Riemann sheet, immediately above and below the cut on the positive real half-line. 

Let's now smoothly cross the positive real half-line, going from the first to the second Riemann sheet.
The value of $F(z)$ above the real axis, on the first sheet, and below it, on the second sheet, is the same by definition:
\beq
F(x+i0^+) = F_{\rm II}(x-i0^+)  \qquad (x>0),
\label{eq:FzeroII}
\eeq
where $F_{\rm II}$  is the function evaluated on the second Riemann sheet.
By using Eqs.\ (\ref{eq:Fzerop})-(\ref{eq:FzeroII}), one gets
\beq
F_{\rm II}(x-i0^+) = F(x-i0^+) + 2 \pi i f(x)
\qquad (x>0).
\label{eq:FIIvsF}
\eeq
Therefore the ``jump" (\ref{eq:Fzerop}) of $F(z)$ evaluated on the two edges of the cut 
(on the first Riemann sheet) is equal to the difference of the values of the function evaluated on the second and first sheet.
By analytically extending formula (\ref{eq:FIIvsF}) one obtains
\beq
F_{\rm II}(z) = F(z) + 2 \pi i f(z).
\qquad \forall z \in \mathbb{C}.
\label{eq:FIIan}
\eeq

\begin{figure}[t]
\begin{center}
\includegraphics[width=8cm]{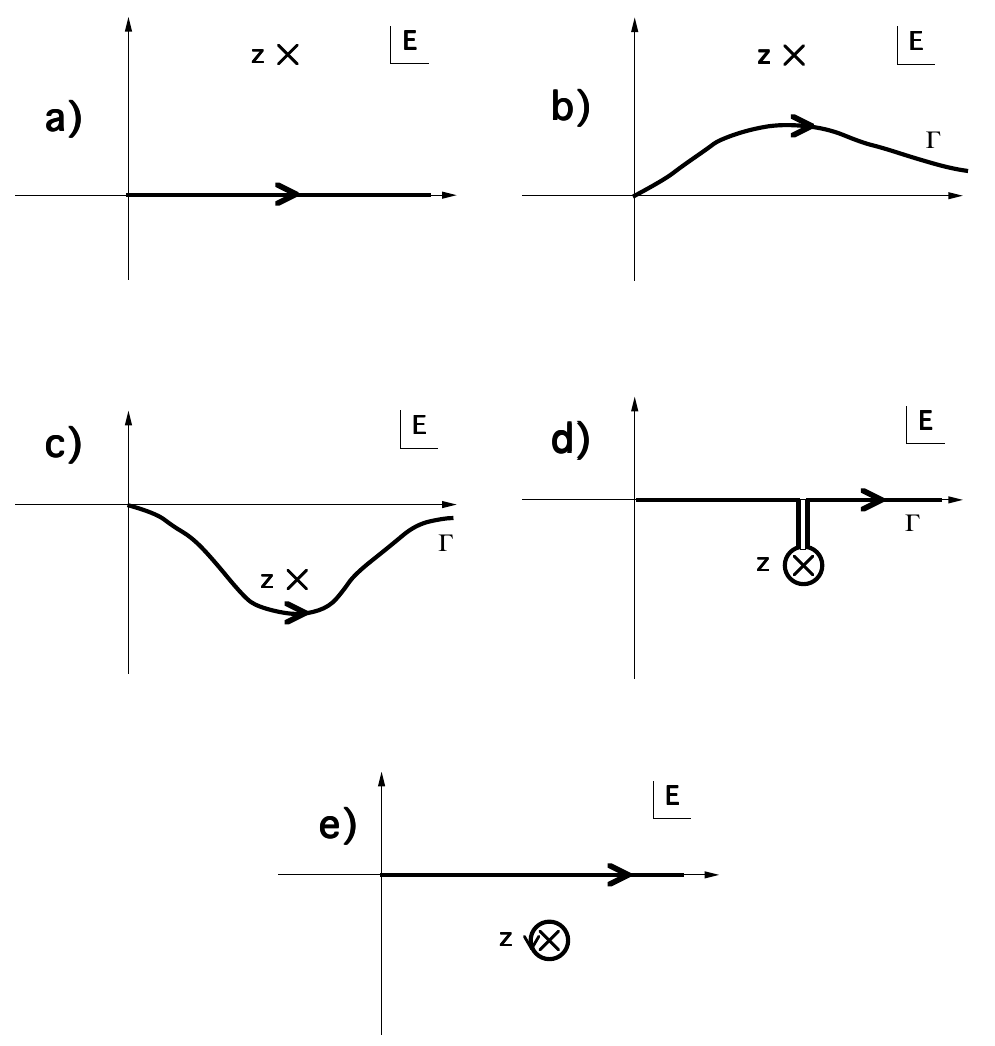}
\end{center}
\caption{Analytic continuation across the cut in the complex $E$-plane.
a) Eq.\ (\ref{eq:Fss}); b) Eq.\ (\ref{eq:Fzz});
c)-e) Eq.\ (\ref{eq:Fexample}). }
\label{fig:Fzcut}
\end{figure}

It is obvious that in the above considerations we are implicitly assuming that analytic continuation is licit. 
Assume now that $F(z)$ in Eq.\ (\ref{eq:Fss}) be defined for $\Im z >0$ and we want to extend it to the region $\Im z <0$. It is easy to see that the definition
\beq
F(z) = \int_\Gamma dE \frac{f(E)}{E-z},    \qquad \mbox{for} \; \Im z>0
\label{eq:Fzz}
\eeq
is equivalent to (\ref{eq:Fss}), as far as the contour $\Gamma$ starts at the origin and reaches $+\infty$ by remaining below $z$.
Notice that $E$ in Eq.\ (\ref{eq:Fzz}) takes complex values and $\Gamma$ can be arbitrarily deformed, as far as its configuration with respect to the singularity $z$ is respected. See Fig.\ \ref{fig:Fzcut}b.
 
The extension to the case $\Im z<0$ is straightforward: when $z$ smoothly crosses the positive real axis, going to the second Riemann sheet, the contour integration in the complex $E$ plane remains below $z$, respecting the position of the singularity. This yields again the result (\ref{eq:FIIan}): the contour is first deformed in order to remain below $z$, then deformed into a small circle, that runs counterclockwise ariound $z$, plus the original contour
\barr
F(x+iy) \stackrel{y<0}{\longrightarrow} F_{\rm II}(x+iy) &=&
\int_\Gamma dE \frac{f(E)}{E-x-iy} \nonumber \\
 &=& \int_0^\infty dE \frac{f(E)}{E-x-iy} + 2\pi i f(x+iy)
\nonumber \\
 &=& F(x+iy) + 2\pi i f(x+iy)
\label{eq:Fexample}
\earr
This is identical to (\ref{eq:FIIan}). 
In this case the difference between $F$ and $F_{\rm II}$ is given by the pole. See Fig.\ \ref{fig:Fzcut}c-e. These beautiful mathematical ideas will be very useful to analyze the behavior of the propagator (\ref{eq:antitran}).

\section{Analytic continuation of the propagator}
 \label{sec-ancontprop}

The function $\As(E)$ in Eqs.\ (\ref{eq:propag}), (\ref{eq:antitran}) has a branching point at $E=\omega_g$, the lower bound of the continuous spectrum of the Hamiltonian $H$, a cut that extends to $E=+\infty$ and no additional sigularities on the first Riemann sheet, while singularities can appear on the second sheet. These important features were studied by Araki {\it et al.} \cite{Araki} and Schwinger \cite{Schwinger} in the 50's\footnote{Those were the golden years of renormalization in quantum field theory.}.

Indeed, $\As(E)$ is defined for $\Im E>0$, so that its Fourier transform, the survival amplitude (\ref{eq:antitran}), converges for $t>0$. When the self-energy function is analytically continued to the second Riemann sheet, the contour must be modified so that its position with respect to the singularity is mantained. 

\begin{figure}[t]
\begin{center}
\includegraphics[width=8cm]{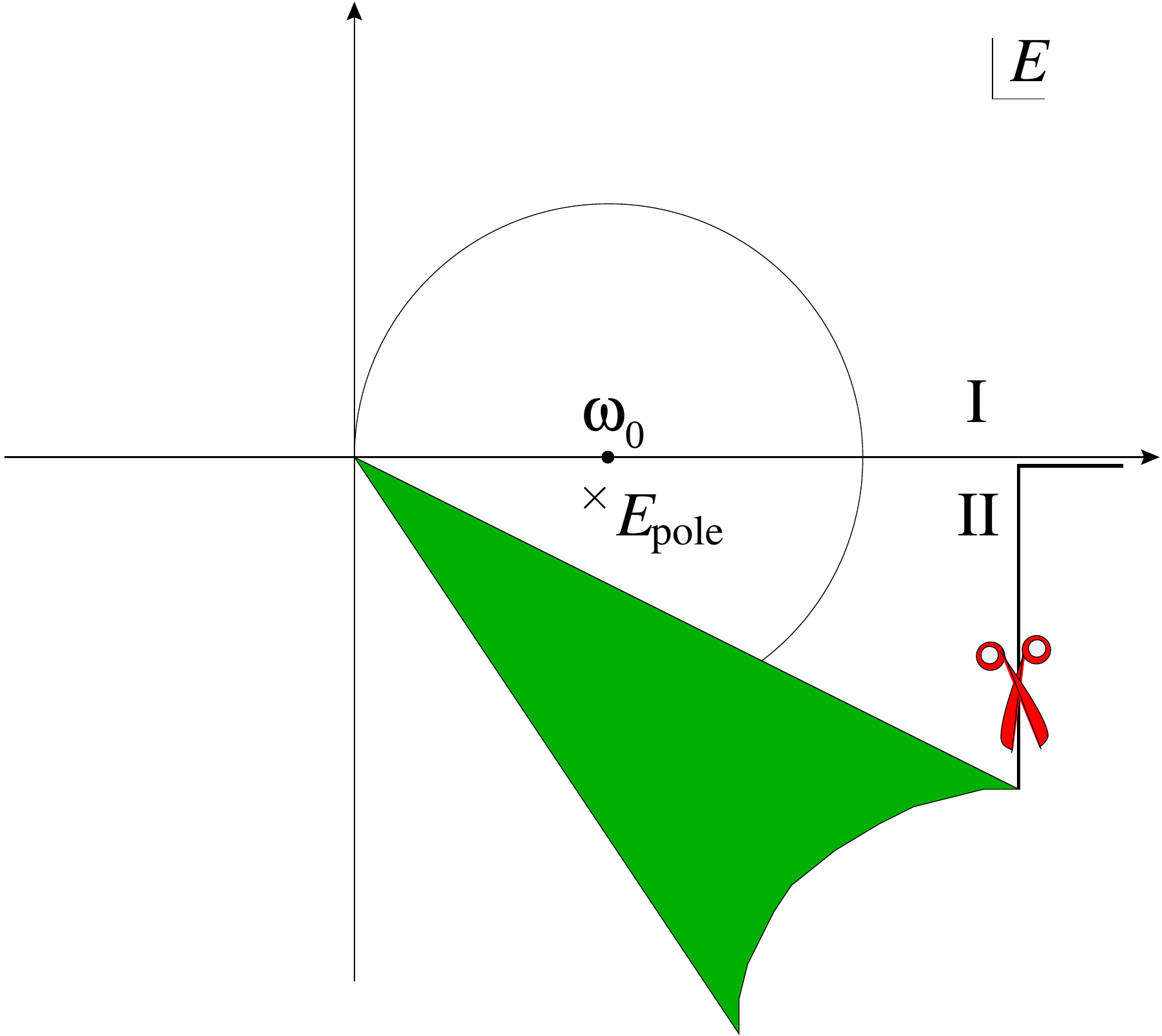}
\end{center}
\caption{The pole $E_{\rm pole}$ on the second Riemann sheet is (coupling constant)$^2$-close to $\omega_0$: see Eqs.\ (\ref{eq:antitran}) and (\ref{eq:1equpol}). We drew the circle of convergence of an asymptotic expansion around $\omega_0$.
The derivation of Eq.\ (\ref{eq:propag}) from Eq.\ (\ref{eq:transf}) requires the definition of the self-energy function (\ref{eq:selfen}). Try and understand which mathematical hypotheses are needed.}
\label{fig:polo}
\end{figure}

The initial state has energy $\omega_0>\omega_g$ and is therefore embedded in the continuous spectrum of $H$. If $|\Sigma(\omega_g)|<\omega_0$ (which happens for sufficiently smooth form factors and small coupling), the resolvent is analytic in the whole complex plane cut along the real axis (continuous spectrum of $H$) \cite{Araki,Schwinger}. On the other hand, there exists a pole $E_{\mathrm{pole}}$ located just below the branch cut in the second Riemann sheet, solution of the equation
\beq
E_{\rm pole}-\omega_0-\Sigma_{{\rm II}}(E_{\rm pole})=0,
\label{eq:1equpol}
\eeq
$\Sigma_{{\rm II}}$ being the determination of the self-energy function in the second sheet. 
Remember that the self-energy function is a ``small" quantity, being proportional to the square of the coupling between level $\ket{+}$ and the continuum: the pole $E_{\mathrm{pole}}$ is therefore very close to $\omega_0$. See Figure \ref{fig:polo}.

The pole has a real and imaginary part
\beq
E_{\rm pole}=\omega_0 + \delta\omega_0-i\gamma/2 ,
\eeq
that can be easily computed by following the mathematical technique outlined in the previous section
\barr
\delta\omega_0&=&\Re \Sigma_{{\rm II}}(E_{\rm pole})\simeq \Re
\Sigma(\omega_0+i0^+)={\rm P}\!\!\int d\omega
\frac{g^2(\omega)}{\omega_0-\omega} ,
\label{eq:2shift}
\\
\gamma&=&-2\Im \Sigma_{{\rm II}}(E_{\rm pole})\simeq -2\Im
\Sigma(\omega_0+i0^+)=2\pi g^2(\omega_0) .
\label{eq:FGR}
\earr
In the above formulas, $\delta\omega_0$ is the energy shift and $\gamma$ the inverse lifetime, according to the Fermi ``golden" rule  \cite{Fermi}. Both quantities are written at second order in the coupling constant.
Check that $\gamma$ is the same quantity that appears in Eq.\ (\ref{eq:goldenrule})\footnote{The derivation of Eqs.\ (\ref{eq:2shift})-(\ref{eq:FGR}) is left as an exercise (a very useful one).
Be careful in deriving $\gamma$ in (\ref{eq:FGR}), you might miss a factor 2. Modern literature (unlike classic literature) is plagued by missing factors 2.
The correct solution is obtained by using the formula
\beq
\lim_{\gamma\rightarrow 0}\frac{\gamma}{E^2+\frac{\gamma^2}{4}}= 2\pi\delta(E), 
\eeq
that is valid because $\gamma$ is a small quantity (second order in the coupling constant), and neglecting 
fourth-order terms in the coupling constant.}.

In conclusion, the survival amplitude (\ref{eq:antitran}) has the general form
\beq
\As(t)=\As_{\rm pole}(t)+\As_{\rm cut}(t),
\eeq
where
\beq
\As_{\rm pole}(t)=\frac{e^{-i(\omega_0+\delta\omega_0)t-\gamma
t/2}}{1-\Sigma'_{{\rm II}}(E_{\rm pole})},
\eeq
is due to the pole contribution (\ref{eq:1equpol}) and
\beq
\label{eq:acut}
\As_{\rm cut}(t)
=\frac{i}{2\pi}\int_{\rm cut}dE\;
\frac{e^{-iEt}}{E-\omega_0-\Sigma(E)},
\eeq
is the branch-cut contribution, as explained in the previous section: see Fig.\ \ref{fig:Fzcut}e).

It is not difficult to see that, if the coupling is small, at intermediate times the pole contribution dominates the evolution and
\beq
P(t)\simeq |\As_{\rm pole}(t)|^2 =  Z e^{-\gamma t} ,\qquad
Z=\left|1-\Sigma'_{{\rm II}}(E_{\rm pole})\right|^{-2} ,
\label{eq:psurv}
\eeq
where $Z$, the intersection of the asymptotic exponential with the $t=0$ axis, is the so-called wave-function renormalization. This explains the behavior sketched in Fig.\ \ref{fig:genevol}. It would be interesting to see  \cite{strev} that the cut contribution (\ref{eq:acut}) \emph{cannot} be neglected at short and long times, where it yields the quadratic Zeno behavior and the power tail, respectively.

\subsection{A few observations}

In order to obtain a purely exponential decay, one can simply neglect the branch cut contribution altogether and retain only the dominant contribution of the pole singularity.
An interesting way to obtain the desired result is to replace the self-energy function with a constant (equal to
its value at the pole) in Eq.\ (\ref{eq:propag}):
\beq
\As(E) \longrightarrow \frac{i}{E-\omega_0-\Sigma_{\rm II}(E_{\rm pole})}
=\frac{i}{E-E_{\rm pole}} \equiv \As^{\mathrm{W}^2}(E) ,
\label{eq:WW0}
\eeq
where we used the pole equation (\ref{eq:1equpol}) in the central equality. This is the celebrated Weisskopf-Wigner approximation \cite{Gamow28} and yields a purely exponential behavior, $\As(t)=\exp(-iE_{\rm pole} t)$, without short- and long-time corrections\footnote{My former teacher M.\ Namiki used to tell me that great physicists know in advance the result they want to get and use mathematics in a ``creative" way to obtain what they need. The older I get, the more I agree.}.

Another nice way to obtain a purely exponential decay is to replace the form factor $g$ in Eq.\ (\ref{eq:contham}) by a constant value, say $\sqrt{\gamma/2\pi}$. This is a useful exercise. (Hint: follow the same strategy as in Sec.\ \ref{sec-QZEnH2}.)

Another important problem is the duration of the non-exponential Zeno region and the onset to the power law. The answer to these questions requires careful evaluation of the cut contribution (\ref{eq:acut}). One finds that the Zeno region is superseded by the exponential decay after a time of the order of the inverse frequency cutoff of the form factor $g$ in the interaction Hamiltonian (\ref{eq:contham}) and the exponential is superseded by a power law after a time of the order of a significant number (say $10^2$) of lifetimes. However, these conclusions are model-dependent and neglect important numerical factors. As a general rule, time evolutions in quantum field theory are a complex problem \cite{Bernardini93} and lead to the inverse Zeno effect \cite{Wilkinson,antiZeno,heraclitus}

\section{Conclusions and apologies.}
\label{sec-concl}

The title of these notes is ``All you ever wanted to know about the quantum Zeno effect in 70 minutes". Admittedly, I lied: my lecture would have lasted 90 minutes, if my chairman had not (very politely) stopped me. However, the title contains a second, more deceitful lie: these notes are by no means \emph{all} you ever wanted to know about the QZE. However, I dont feel guilty about the second (white) lie. The main purpose of a lecture is not to explain everything; rather, it is to make students curious, so that they can go and deepen the subject. This contains, in embryo, what we nowadays call curiosity-driven research. If I managed to get my students interested, my lecture was successful.

\section*{Acknowledgments}
I would like to thank D.\ Chru\'sci\'nski, A.\ Jamio\l kowski and M.\ Michalski for the kind invitation to lecture at the 44th Symposium on Mathematical Physics ``New Developments in the Theory of Open Quantum Systems", held in Toru\'n, Poland, in June 20-24, 2012. 
Special thanks to B.\ Bylicka and F.\ Pepe for suggestions and comments and to P.\ Facchi and H.\ Nakazato for many early conversations on the quantum Zeno phenomenon. Many thanks to A.\ Takahashi for suggesting the title.
This work was partially supported by PRIN 2010LLKJBX on ``Collective quantum phenomena: from strongly correlated systems to quantum simulators".


\begin{thebibliography}{91}
\bibliographystyle{plain}

\bibitem{strev}
H. Nakazato, M. Namiki and S. Pascazio, Int. J. Mod. Phys. B \textbf{10}, 247 (1996).

\bibitem{zenoreview}
P. Facchi and S. Pascazio, Progress in Optics, edited by E.\ Wolf (Elsevier, Amsterdam) \textbf{42}, 147 (2001);
J. Phys. A: Math. Theor. \textbf{41}, 493001 (2008).

\bibitem{Misra}
B. Misra and E. C. G. Sudarshan, J. Math. Phys. \textbf{18}, 756 (1977).

\bibitem{KK}
K. Koshino and A. Shimizu, Phys. Rep. \textbf{412}, 191 (2005). 

\bibitem{aristotle}
Aristotle, Physics 4 239b10.

\bibitem{video}
\url{http://tv.umk.pl/#channel=3,movie=1748}

\bibitem{waseda01}
P. Facchi and S. Pascazio,
``Unstable systems and quantum Zeno phenomena in quantum field theory"
Quantum Probability and White Noise Analysis \textbf{XVII}, 222 (2003). [quant-ph/0202127]

\bibitem{opticalpotential}
E. Fermi, Ricerca Scientifica \textbf{7}, 13 (1936);
E. Fermi, Nuovo Cimento \textbf{11}, 407 (1954);
H. Feshbach, C.E. Porter and V.F. Weisskopf, Phys. Rev. \textbf{96}, 448 (1954).

\bibitem{fermizinn}
E. Fermi and W. H. Zinn, Phys. Soc. Cambridge Conf. Rep. 92, Chicago, 1947.
[See Enrico Fermi, Collected Papers, ed. E. Segr\`e (University of Chicago Press, 1962), Paper no. 220.]

\bibitem{GKSL}
V. Gorini, A. Kossakowski and E. C. G. Sudarshan ,
J. Math. Phys. {\bf 17}, 821 (1976);
G. Lindblad, Comm. Math. Phys. {\bf 48} 119 (1976)

\bibitem{Alicki} 
R. Alicki and K. Lendi, {\it Quantum Dynamical
Semigroups and Applications} (Springer, Berlin, 1987).

\bibitem{Weiss}
U. Weiss, {\it Quantum Dissipative Systems}, (World
Scientific, Singapore, 2000).

\bibitem{Breuer}  
H.-P. Breuer  and F. Petruccione, {\em The Theory of Open Quantum Systems} (Oxford Univ. Press, Oxford, 2007).

\bibitem{colhepp}
H. Nakazato and  S. Pascazio, Phys. Rev. A \textbf{48}, 1066 (1993). 

\bibitem{measdiff}
P. Facchi, S. Pascazio and A. Scardicchio,
Phys. Rev. Lett. \textbf{83}, 61 (1999). 

\bibitem{Fermi}
E. Fermi, Rev. Mod. Phys. \textbf{4} 87 (1932);
\textit{Nuclear Physics} (University of Chicago, Chicago, 1950) p.\ 136, 142, 148;
\textit{Notes on Quantum Mechanics. A Course Given at the University of Chicago in 1954}, edited by E. Segr\`e (University of Chicago, Chicago, 1954) Lec.\ 23.

\bibitem{heraclitus} P. Facchi, H. Nakazato and S. Pascazio,
Phys. Rev. Lett. \textbf{86}, 2699 (2001).

\bibitem{von}
J. von Neumann,
\textit{Die Mathematische Grundlagen der Quantenmechanik} (Springer,
Berlin, 1932). [English translation by E. T. Beyer,
\textit{Mathematical Foundation of Quantum Mechanics} (Princeton
University Press, Princeton, 1955)].

\bibitem{pollQM}
M. Schlosshauer, J. Kofler and A. Zeilinger,
``A Snapshot of Foundational Attitudes Toward Quantum Mechanics",
Stud. Hist. Phil. Mod. Phys. \textbf{44}, 222 (2013)
([quant-ph] arXiv:1301.1069).

\bibitem{WZ83}
 J.A. Wheeler and W.H. Zurek, eds, \textit{Quantum Theory and Measurement} (Princeton University 
Press, 1983). 

\bibitem{NP93}
M. Namiki and  S. Pascazio,
Phys. Rept. \textbf{232}, 301 (1993).

\bibitem{Schulman98}
L.S. Schulman, Phys. Rev. A {\bf 57}, 1509 (1998).

\bibitem{Araki}
H. Araki, Y. Munakata, M. Kawaguchi and T. Goto, Progr. Theor. Phys. {\bf 17} (1957) 419.

\bibitem{Schwinger}
J. Schwinger, Ann. Phys. {\bf 9} (1960) 169.

\bibitem{Gamow28}
G. Gamow, Z. Phys. {\bf 51}, 204 (1928); 
V. Weisskopf and E.P. Wigner, Z. Phys. {\bf 63}, 54 (1930);  {\bf 65}, 18 (1930); 
G. Breit and E.P. Wigner, Phys. Rev. {\bf 49}, 519 (1936).

\bibitem{Bernardini93}
C. Bernardini, L. Maiani and M. Testa, Phys. Rev. Lett. {\bf 71}, 2687 (1993); 
L. Maiani and M. Testa, Ann. Phys. (NY) {\bf 263}, 353 (1998); 
P. Facchi and S. Pascazio, Phys. Lett. A {\bf 241}, 139 (1998); Physica A \textbf{271}, 133 (1999);
I. Joichi, Sh. Matsumoto, and M. Yoshimura, Phys. Rev. D {\bf 58}, 045004 (1998);
R.F. Alvarez-Estrada and J.L. S\'anchez-G\'omez, Phys. Lett. A {\bf 253}, 252 (1999);
A.D. Panov, Physica A {\bf 287}, 193 (2000);
I. Antoniou, E. Karpov, G. Pronko and E. Yarevsky, Phys. Rev. A \textbf{63} 062110 (2001);
M. Gadella and G. P. Pronko, Fortschr. Phys. \textbf{59}, 795 (2011)


\bibitem{Wilkinson}
S.R. Wilkinson, C.F. Bharucha, M.C. Fischer, K.W. Madison, P.R. Morrow, Q. Niu, B. Sundaramand M.G. Raizen, Nature \textbf{387}, 575 (1997);
M.C. Fischer, B. Guti\'errez-Medina and M.G. Raizen Phys. Rev. Lett. \textbf{87}, 040402 (2001)

\bibitem{antiZeno}
A.M. Lane, Phys. Lett. A \textbf{99} 359 (1983);
W.C. Schieve, L.P. Horwitz and J. Levitan, Phys. Lett. A \textbf{136}, 264 (1989)
A.G. Kofman and G Kurizki, Nature \textbf{405}, 546 (2000);
J. \v{R}eh\'a\v{c}ek, J. Pe\v{r}ina, P. Facchi, S. Pascazio, and L. Mi\v{s}ta,
Phys. Rev. A \textbf{62}, 013804 (2000).  

\end{thebibliography}
\end{document}